\documentclass[preprint,showpacs]{revtex4}
\usepackage{bm}% bold math

\ifx\pdfoutput\undefined
\usepackage{graphicx}
\else
\usepackage[pdftex]{graphicx}
\usepackage{epstopdf}
\fi

\usepackage[center]{subfigure}

\begin{document}

\title{Renormalization group approach to multiscale modelling in materials science}

\author{Nigel Goldenfeld}
%\email{nigel@uiuc.edu}
\affiliation{Department of Physics, University of Illinois at
Urbana-Champaign, 1110 West Green Street, Urbana, Illinois, 61801-3080.}
\author{Badrinarayan P. Athreya and Jonathan A. Dantzig}
\affiliation{Department of Mechanical and Industrial Engineering, 1206
West Green Street, Urbana, IL 61801.}
%\date{}

\begin{abstract}

Dendritic growth, and the formation of material microstructure in
general, necessarily involves a wide range of length scales from the
atomic up to sample dimensions.  The phase field approach of Langer,
enhanced by optimal asymptotic methods and adaptive mesh refinement,
copes with this range of scales, and provides an effective way to move
phase boundaries.  However, it fails to preserve memory of the
underlying crystallographic anisotropy, and thus is ill-suited for
problems involving defects or elasticity.  The phase field crystal
(PFC) equation--- a conserving analogue of the Hohenberg-Swift equation
---is a phase field equation with periodic solutions that
represent the atomic density.  It can natively model elasticity, the
formation of solid phases, and accurately reproduces the nonequilibrium
dynamics of phase transitions in real materials. However, the PFC
models matter at the atomic scale, rendering it unsuitable for coping
with the range of length scales in problems of serious interest.  Here,
we show that a computationally-efficient multiscale approach to the PFC
can be developed systematically by using the renormalization group or
equivalent techniques to derive appropriate coarse-grained coupled
phase and amplitude equations, which are suitable for solution by
adaptive mesh refinement algorithms.

\end{abstract}

%Relevant Pacs numbers (up to four allowed, in order of relevance)
%
%05.70.Ln Nonequilibrium and irreversible thermodynamics
%81.16.Rf Nanoscale pattern formation
%05.10.Cc Renormalization group methods
%81.15.Aa Theory and models of film growth
%61.72.Cc Kinetics of defect formation and annealing
%64.60.Cn Order-disorder transformations; statistical mechanics of model systems

\maketitle

\section{Introduction}

During the last thirty years or so, the field of computational
materials science has emerged as a flourishing sub-discipline of
condensed matter physics.  It is now relatively straightforward to
compute realistic-looking materials microstructures for a variety of
processing conditions, and, with enough computer power, to begin to
make quantitative predictions about phase diagrams, morphological phase
diagrams, growth rates and other probes of the kinetics of phase
transitions.  But this was not so, nor even obviously within the realms
of possibility, during the 1970's when Jim Langer, Pierre Hohenberg and
others initiated studies of phase transitions kinetics and
instabilities in spatially-extended systems, whose fruitful union in
the 1980's gave birth to the field of pattern formation as practiced
today.  In those early days, the key problems were to understand the
nature of the instabilities around uniform or similarity solution
states, and to capture correctly the intricate non-local and retarded
feedback between interface dynamics and the (typically) diffusion
fields around them. To paraphrase John Archibald Wheeler's epigram on
general relativity: solidification fronts tell heat how to flow;
heat tells solidification fronts how to move.

Following a variety of innovations from Langer and collaborators,
especially the introduction of phase field models\cite{Langer, COLL85},
and their practical implementation with improved
asymptotics\cite{Karma} and adaptive mesh refinement\cite{PGD}, these
difficulties are largely resolved, certainly in principle, and to a
significant extent in practice\cite{Warren}.  Three dimensional
structures, including fluid flow effects, can now be simulated on
desktop computers\cite{Jeong2,Jeong1}. The key conceptual difficulty
that remains as a challenge to this day is the huge range of length and
time scales encountered in pattern formation processes.  For example,
in solidification, one encounters the capillary length on the scale of
$10^{-8} m$ and the dendrite tip radius on a scale ranging from tens of
microns to millimetres, depending on the undercooling.  The latent heat
around a solidification front extends to a distance known as the
diffusion length, which can be $10^{-4}m$ or larger; and finally, in a
real processing experiment, there will be the system size itself, many
orders of magnitude larger still.

The problem of \lq\lq bridging the length and time scales" between
atomic and sample dimensions is the focus of the present article.
Despite much activity to address the scale-up
problem\cite{Phillipsbook, VVED04}, including quasi-continuum
methods\cite{Tadmor, Shenoy, Ortiz, Miller}, the heterogeneous
multiscale method\cite{Weinan1, Weinan2}, multi-scale molecular
dynamics\cite{Rudd, Kaxiras, Robbins, CURT02} and multigrid
variants\cite{Fish}, most existing work is currently limited to
crystalline materials with a few isolated defects\cite{Weinan3}.  We
mention as an exception a promising approach\cite{warren03} based on
the phase field approach, due originally to Langer (but not published
for many years)\cite{Langer}.

The approach advocated here introduces the atomic level of description
into phase field approaches using a recently developed formalism known
as the phase field crystal (PFC)\cite{ekhg02, eg04}.  The PFC describes
atom crystalline structure as a periodic density wave, and posits a
natural equation of motion for the density field.  Having the character
of a continuum partial differential equation, it can be coarse-grained
using renormalization group (RG)\cite{GMOL, NGbook} and related methods
(see, e.g. \cite{BOWM98}), developed for the quite different problem of
analyzing hydrodynamic instabilities\cite{Cross} in spatially-extended
dynamical systems\cite{CGO2, Graham, Nozaki, Sasa, Shiwa, CN, PN, NPL}.
The coarse-grained counterparts of the density field are its amplitude
and phase, for which we derive equations of motion.  Once known, the
density field can then be reconstructed. In this article, we sketch the
derivation of effective equations at the mesoscale, and show that the
numerical solution of the renormalization group equations generates
solutions that are virtually indistinguishable from brute force
solutions of the PFC equations of motion.

A further aspect of our work is that computational efficiency is in
practice best achieved by using adaptive mesh refinement.  The basic
idea is that for any field whose spatial variation is essentially
uniform everywhere, with localized regions of rapid variation, it makes
no sense to use a uniform mesh in numerical calculations.  A coarse
mesh can be used in places where the field is spatially uniform, and a
coarser one is used in the transition zones. In the context of phase
field models, we have developed techniques to implement these ideas in
an efficient manner, and demonstrated that the computational complexity
of this sort of algorithm scales not with the system's volume, but
with the surface area of solidification front (transition zones)---a
considerable saving for large systems\cite{PGD}.  The renormalization
group equations derived below have solutions with the desired
character: thus we intend in future work to develop adaptive mesh
refinement codes for solving these equations.  This topic is briefly
discussed at the end of this article.

\section{The Phase Field Crystal Model}

The phase field crystal model for a single component system describes
the space-time behavior of the density $\rho(\vec{x},t)$ and is
capable\cite{ekhg02, eg04} of capturing realistic aspects of materials
dynamics, including grain growth, ductile fracture, epitaxial growth,
solidification processes, and reconstructive phase transitions.
By construction, the stationary states of $\rho(\vec{x})$ are periodic,
and distortions of the density field by external perturbations applied
to boundaries or by defects, for example, result in a raising of the
free energy in accord with Hooke's law, with higher order terms
representing non-linear elasticity.  This feature makes the PFC an
ideal tool with which to explore nanoscale strain effects and their
influence upscale to the continuum.

Let $F\{\rho\}$ denote the coarse-grained free energy functional whose
minima correspond to the equilibrium (lattice) state of a
$d$-dimensional system, and whose corresponding chemical potential
gradient drives the dynamics of $\rho$.  For single component,
two-dimensional systems with a hexagonal lattice state, the appropriate form
of $F$ is that originally due to
Brazovskii\cite{Brazovskii}:
\begin{equation}
F\{\rho({\bf x})\}=\int d^d\vec{x}\left[ \rho\left(\alpha\Delta T +
\lambda\left(q_o^2+\nabla^2\right)^2\right)\rho/2+u\rho^4/4\right]
\label{eq:Free_energy}
\end{equation}

\noindent where $\alpha$, $\lambda$, $q_o$ and $u$ can be related to
material properties\cite{ekhg02,eg04}, and $\Delta T$ denotes the
temperature difference from some reference high temperature.  For
convenience it is useful to rewrite this free energy in dimensionless
units, i.e., $\vec{x} \equiv \vec{r}q_o$, $\psi \equiv \rho
\sqrt{u/\lambda q_o^4}$, $r\equiv a\Delta T/\lambda q_o^4$ and
$F\rightarrow F u/\lambda^2q_o^{8-d}$.  We also use dimensionless time
units $t$, related to the physical time $\tilde t$ in terms of the
phenomenological mobility $\Gamma$ by $t \equiv \Gamma\lambda q_o^6
\tilde t$.  In these units, the equation of continuity for the density
becomes
\begin{equation}
\label{eq:dyn}
\partial \psi/\partial t = \nabla^2\left(\left[r + (1 + \nabla^2)^2\right]
\psi +\psi^3\right)+\zeta.
\end{equation}

\noindent The conserved Gaussian noise, required by the
fluctuation-dissipation theorem to satisfy $\langle
\zeta(\vec{r}_1,t_1)\zeta(\vec{r}_2,\tau_2) \rangle = -{\cal E}
\nabla^2\delta(\vec{r}_1-\vec{r}_2)\delta(\tau_1-\tau_2)$ with ${\cal
E} \equiv uk_BTq_o^{d-4}/\lambda^2$, will not generally be important for
describing phase transition kinetics, and so will henceforth be
neglected here.

Elder and Grant\cite{eg04} have studied the mean field phase diagram of
the PFC equation (\ref{eq:dyn}) in the one mode approximation, that is
valid in the limit of small $r$, and represented in the plane of
dimensionless temperature, $r$, and dimensionless average density,
$\bar{\psi}$. There are three possible equilibrium solutions: a
`liquid', $\psi_C = \bar{\psi}$, a two-dimensional `crystal' with
triangular symmetry, $\psi_T = A_T \left(\cos(q_Tx)\cos(q_Ty/\sqrt{3})-
\cos(2q_Ty/\sqrt(3))/2\right)+\bar{\psi}$, and a smectic phase which
will be ignored for present purposes.  The triangular lattice can
exhibit defect structures during the relaxation to equilibrium, thus
capturing the kinetics of phase transformations.

\section{Coarse-graining of the phase field crystal equations}

The dynamics of the slowly-varying amplitude and phase describes
fluctuations about a given set of lattice vectors, but must be
covariant with respect to rotations of those lattice vectors, so that
we can properly describe polycrystalline materials with arbitrarily
oriented grains.  A similar situation arises in describing amplitude
and phase variations of convection rolls, and in the context of the
model Swift-Hohenberg \cite{SH} equations, the form of the governing
equations was originally proposed by Gunaratne et al. \cite{gunaratne},
and derived systematically from the RG formalism of Chen et
al.\cite{CGO2} by Graham\cite{Graham} (see also ref. \cite{Nozaki}).

The details of the corresponding calculations for the PFC will be given
elsewhere, but the results can be quickly obtained by considering the
linearized equation governing small variations $\widetilde{\psi}$ about
the {\it constant\/} solution $\bar\psi$:
\begin{equation}
\label{lineg}
\frac{\partial \widetilde{\psi}}{\partial t}  =  \nabla^2 \left[ \left\{r + (1 +
\nabla^2)^2\right\} + 3\bar{\psi}^2\right]\widetilde{\psi}
\end{equation}

We consider a one-mode triangular perturbation with wavenumber $q_t$,
given by $\widetilde{\psi} = \exp(\omega
t)\left[\cos(q_tx)\cos(q_ty/\sqrt{3})-\cos(2q_ty/\sqrt{3})/2\right]$.
Substitution into (\ref{lineg}) yields the following dispersion
relation:
\begin{equation}
\label{ddrtriang} \omega =
-\frac{4}{3}q_t^2\left[r+3\bar{\psi}^2+\left(1-\frac{4}{3}q_t^2\right)^2\right].
\end{equation}

The fastest growing mode has $q_t = \pm\sqrt{3}/2$ for this
perturbation, which, as was shown by Elder and Grant\cite{eg04} working
in the one-mode approximation, is the mode which minimizes the free
energy functional $\mathcal{F}[\rho]$.  Thus, the uniform phase becomes
unstable to hexagonal perturbations when  $|\bar{\psi}| < \sqrt{-r/3}$.

The triangular phase solution can be written in more general form as
\begin{equation}
\label{1modetriang}
\psi({\vec{x}})=\sum_{j}A_{j}(t)\exp(i{\vec{k_{\it{j}}}\cdot\vec{x}})
+ \bar{\psi},
\end{equation}
where ${\vec{k_{1}}} = k_0 ({-\vec{\mbox{i}}\sqrt{3}/2} -
{\vec{\mbox{j}}/2})$, ${\vec{k_{2}}} = k_0{{\vec{\mbox{j}}}}$ and
${\vec{k_{3}}} = k_0({\vec{\mbox{i}}\sqrt{3}/2} - {\vec{\mbox{j}}/2})$
are the reciprocal lattice vectors, $\vec{\mbox{i}}$ and
$\vec{\mbox{j}}$ are unit vectors in the $x$- and $y$-directions, $k_0$
is given by $q_t$ and $A_{j}(t)\equiv A_{j}^0\exp(\omega_{j}t)$.

Substituting into (\ref{lineg}) and applying orthogonality
conditions, we readily obtain
\begin{equation}
\omega_{j} = -|{\vec{k_{\it{j}}}}|^2\left[-\Gamma+
\left\{1-|{\vec{k_{\it{j}}}}|^2\right\}^2
\right],
\end{equation}
where $\Gamma \equiv -(r+3\bar{\psi}^2)$, and we have transformed to
units where $k_0=1$. In order to obtain the spatial component of the
amplitude equation we study the effect of applying small spatial
modulations to the amplitude, i.e.
\begin{equation}
A_{j}(t) \longmapsto A_{j}(x,y,t) =
A_{j}^0\exp(\omega_{j}({\vec{Q}})t)\exp(i{\vec{Q}\cdot\vec{x}}),
\end{equation}
where ${\vec{Q}}=Q_x\;{\vec{i}}+Q_y\;{\vec{j}}$ is the perturbation
vector. The exponent controlling the growth rate along each lattice
basis vector is given by
\begin{equation}
\label{ddr_pert}
\omega_{j}({\vec{Q}}) = |{\vec{Q}}+{\vec{k_{\it{j}}}}|^2\left[\Gamma -
\left\{1-|{\vec{Q}}+{\vec{k_{\it{j}}}}|^2\right\}^2
\right].
\end{equation}
Replacing Fourier variables by gradient
operators, we obtain
\begin{equation}
\label{prim_oper}
|{\vec{Q}}+{\vec{k_{\it{j}}}}|^2 \equiv -{\vec{\nabla}}^2 -
2i{\vec{k_{\it{j}}}}\cdot{\vec{\nabla}} + 1.
\end{equation}
Putting together Eq.~(\ref{ddr_pert}) and Eq.~(\ref{prim_oper}), we
find that
space-time amplitude variations along each basis are governed by:
\begin{equation}
\label{amp_vars}
\frac{\partial A_{j}}{\partial t} = \left[1 - {\vec{\nabla}}^2 -
2i{\vec{k_{\it{j}}}}\cdot{\vec{\nabla}}\right]
\left[\Gamma-
\left\{{\vec{\nabla}}^2 +
2i{\vec{k_{\it{j}}}}\cdot{\vec{\nabla}}\right\}^2\right]A_j.
\end{equation}
The manifestly rotationally covariant operator on the right hand side of
(\ref{amp_vars}) will be denoted by $\widetilde{\mathcal{L}}_{j}$.

The non-linear component of the amplitude equation is obtained by
renormalizing the secular terms in the perturbation series about the
periodic state\cite{CGO2, Graham, Nozaki}.  Expanding about the
one-mode triangular phase solution, scaling $\psi$ by the factor
$\sqrt{-r}$, and renaming the new variable $\psi$,yields the PFC
equation in the form
\begin{equation}
\label{eg_scaled}
\frac{\partial \psi}{\partial t}  = \vec{{\nabla}}^2(1 + \vec{{\nabla}}^2)^2\psi
+ \epsilon\vec{{\nabla}}^2(\psi^3- \psi),
\end{equation}
where $\epsilon \equiv -r$. Expanding $\psi$ as $\psi = \psi_0 +
\epsilon\psi_1 + \epsilon^2\psi_2 + \mathcal{O}(\epsilon^3)$ and
substituting in Eq.~(\ref{eg_scaled}), using Eq.~(\ref{1modetriang})
for $\psi_0$, we obtain at $\mathcal{O}(\epsilon)$,
\begin{equation}
\label{eg_eik}
\frac{\partial \psi_1}{\partial t}  - \vec{{\nabla}}^2(1 + \vec{{\nabla}}^2)^2\psi_1
= \vec{{\nabla}}^2(\psi_0^3- \psi_0).
\end{equation}
The secular terms on the right hand side can be identified from the
expansion
\begin{eqnarray}
\label{secterms}
\vec{{\nabla}}^2(\psi_0^3- \psi_0) =  (1-3\bar{\psi}^2)\sum_{j}A_{j}
\exp(i{\vec{k_{\it{j}}}\cdot\vec{x}})\nonumber\\
 -3A_1\left(|A_1|^2+2|A_2|^2+2|A_3|^2\right)\exp(i{\vec{k}_{1}\cdot\vec{x}})\nonumber\\
 -3A_2\left(2|A_1|^2+|A_2|^2+2|A_3|^2\right)\exp(i{\vec{k}_{2}\cdot\vec{x}})\nonumber\\
 -3A_3\left(2|A_1|^2+2|A_2|^2+|A_3|^2\right)\exp(i{\vec{k}_{3}\cdot\vec{x}})\nonumber\\
 -6A_2^*A_3^*\bar{\psi}\exp(i{\vec{k}_{1}\cdot\vec{x}})
 -6A_1^*A_3^*\bar{\psi}\exp(i{\vec{k}_{2}\cdot\vec{x}})\nonumber\\
 -6A_1^*A_2^*\bar{\psi}\exp(i{\vec{k}_{3}\cdot\vec{x}}).
\end{eqnarray}

Aligning the secular terms in Eq.~(\ref{secterms}) along each basis,
and scaling back to the original variables, we obtain the amplitude
equations as
\begin{equation}
\label{rgeqn}
\frac{\partial A_1}{\partial t} =
\widetilde{\mathcal{L}}_{1}A -
3A_1\left(|A_1|^2+2|A_2|^2+2|A_3|^2\right) - 6\bar{\psi}A_2^*A_3^*.
\end{equation}
together with the appropriate permutations for $A_2$ and $A_3$.

\begin{figure}[htbp]
\begin{center}
\subfigure[\label{eg8} $t=56.04$]
{\includegraphics[height=2.2cm,angle=0]{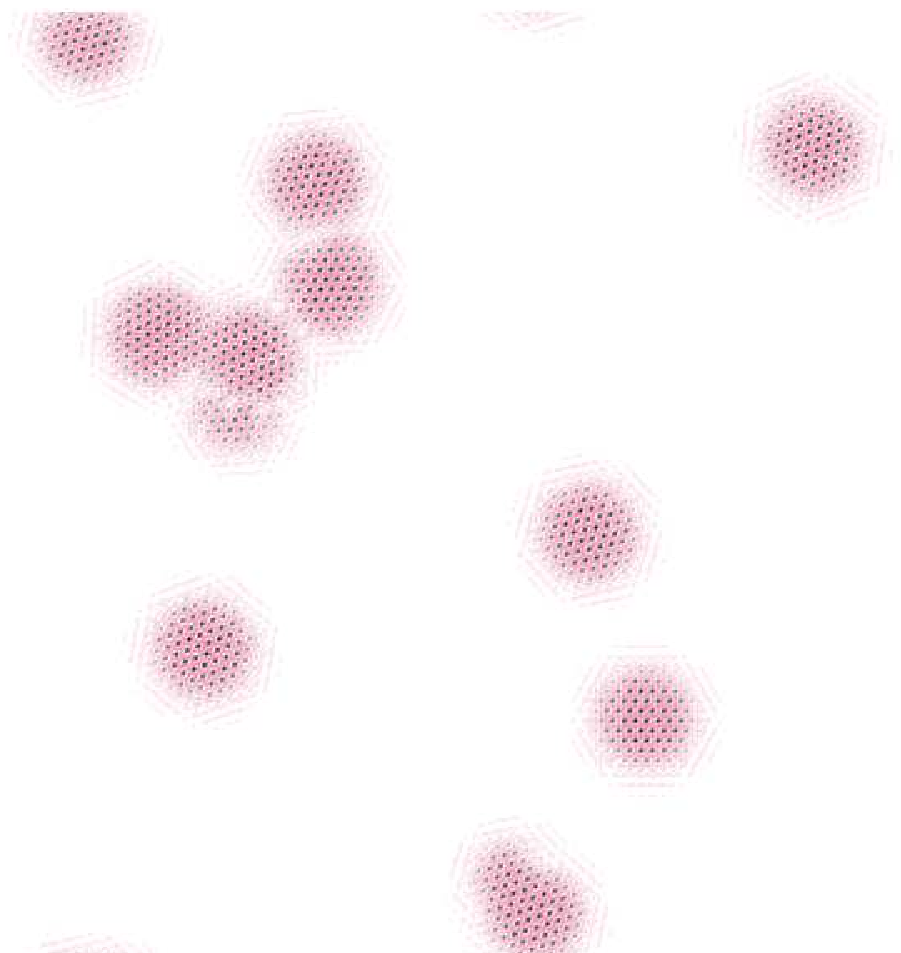}}
\hfill
\subfigure[\label{eg24} $t=184.04$]
{\includegraphics[height=2.2cm,angle=0]{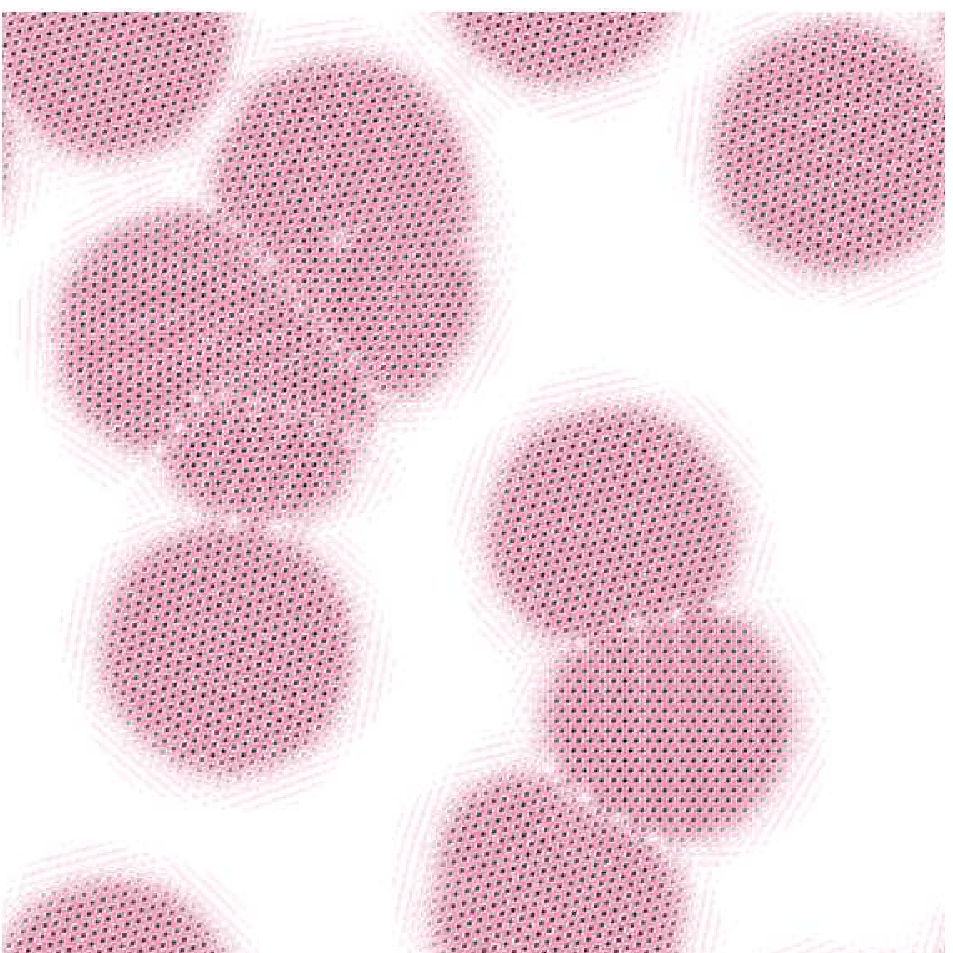}}
\hfill
\subfigure[\label{eg64} $t=720.04$]
{\includegraphics[height=2.2cm,angle=0]{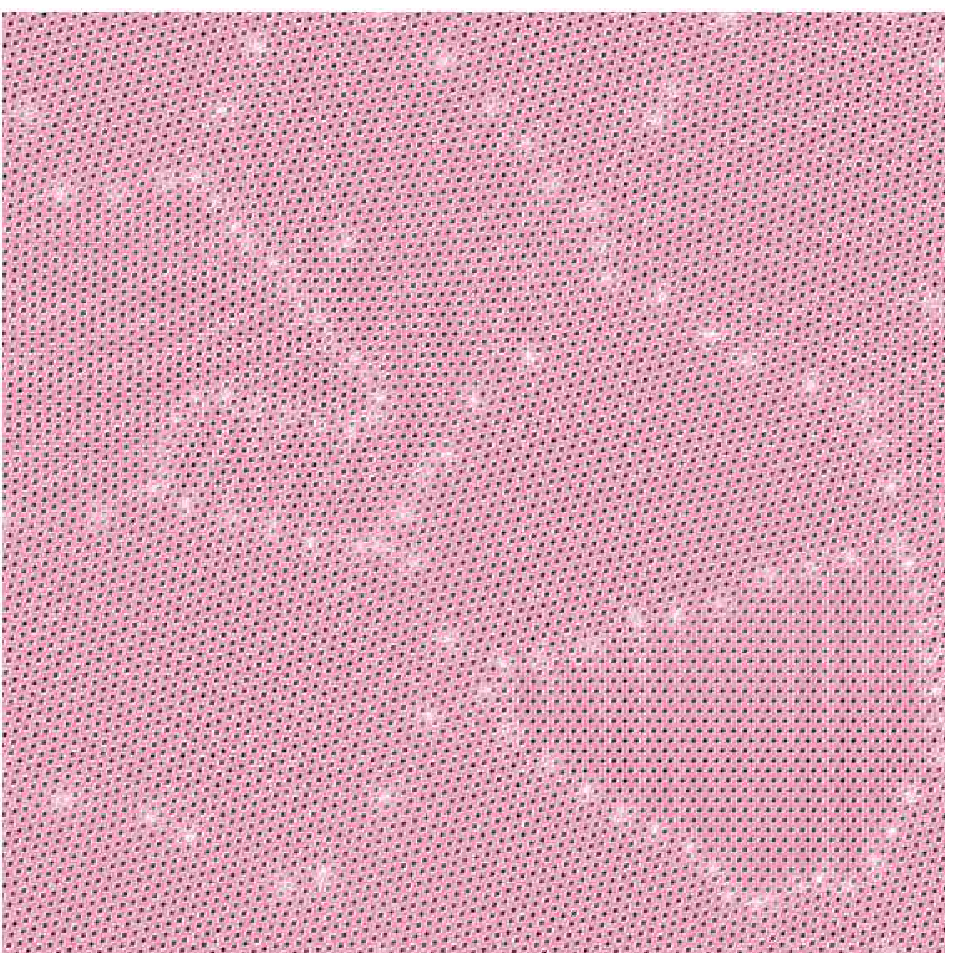}}
\hfill
\subfigure[\label{egrg8} $t=56.04$]
{\includegraphics[height=2.2cm,angle=0]{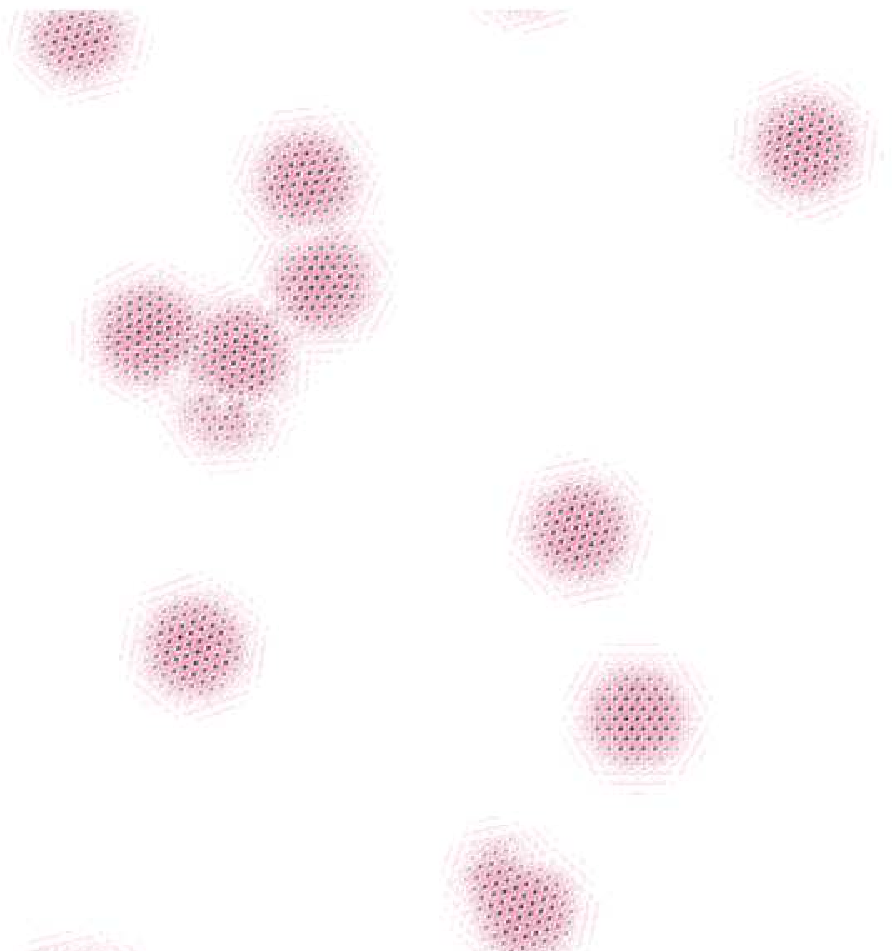}}
\hfill
\subfigure[\label{egrg24} $t=184.04$]
{\includegraphics[height=2.2cm,angle=0]{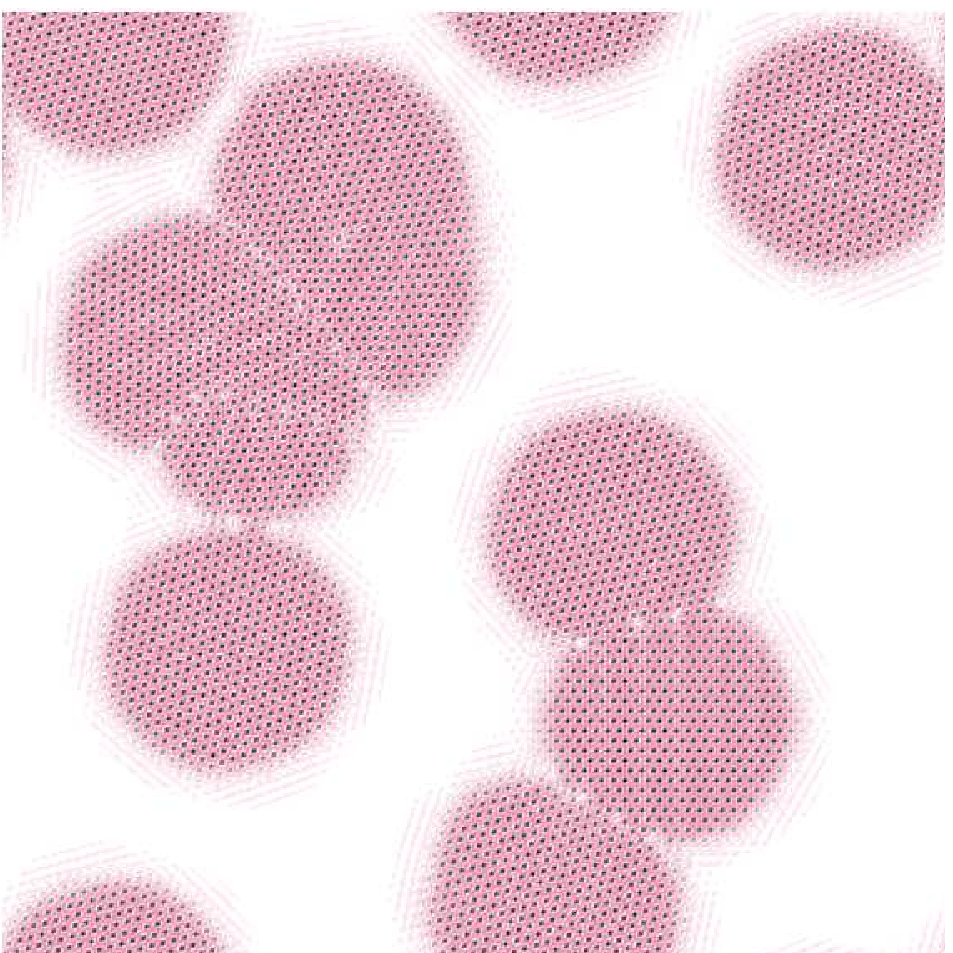}}
\hfill
\subfigure[\label{egrg64} $t=720.04$]
{\includegraphics[height=2.2cm,angle=0]{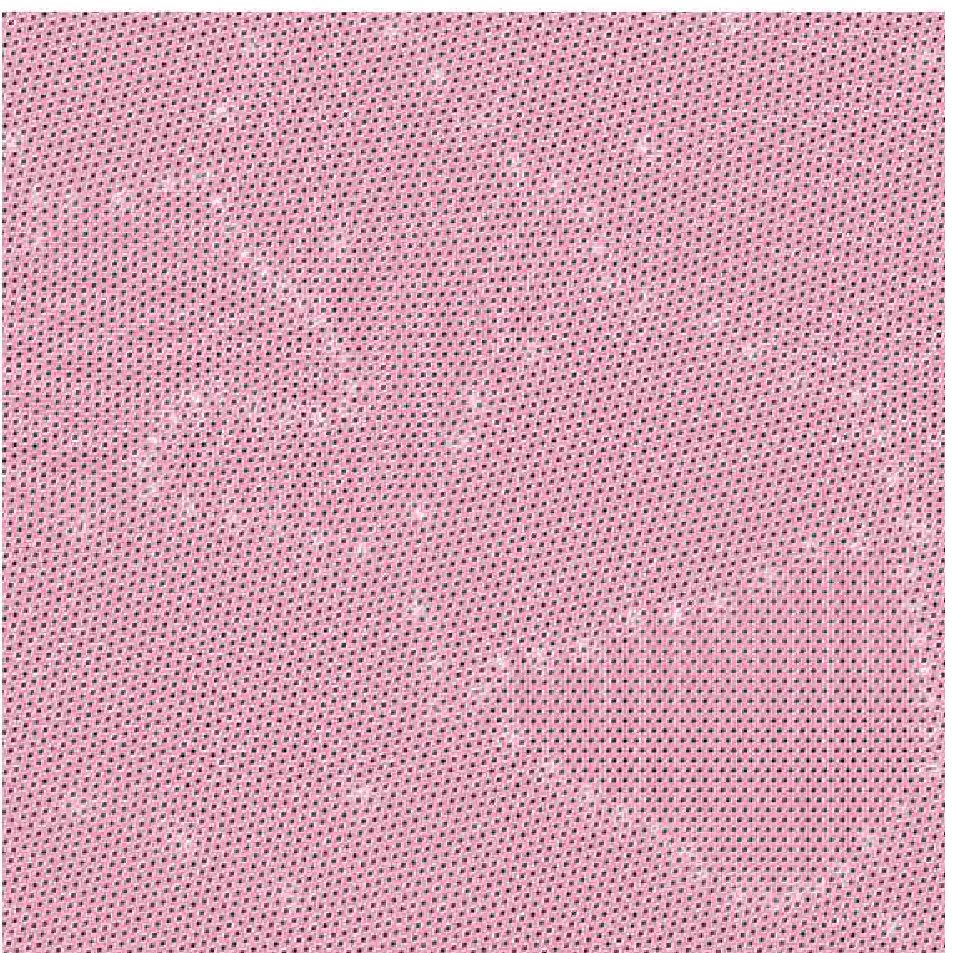}}
\caption{\label{fig:comparison} Comparison of
heterogeneous nucleation and growth in the PFC equation
Eq.~(\ref{eq:dyn}) (panels (a)-(c)) and its RG-generated mesoscale
counterpart Eq.~(\ref{rgeqn}) (panels (d)-(f)).  The order parameter is
shown at the times indicated starting from the same initial condition
with $\bar{\psi} = 0.285$ and $r=-0.25$.}
\end{center}
\end{figure}

Fig.~(\ref{fig:comparison}) shows the time evolution for the nucleation
and growth of a two-dimensional film, calculated using the PFC equation
and its RG-generated mesoscale counterpart.  Starting from the same
initial condition of randomly-oriented seeds, crystalline domains grow,
colliding to form a polycrystalline microstructure.  The solutions from
the two different computational algorithms are essentially
indistinguishable.  Without the PFC formulation, it would not have been
possible to capture successfully the formation of a polycrystalline
material, with grain boundaries and other defects.  We conclude that it
is indeed possible to compute large scale microstructure from effective
equations at the mesoscale.

\section{Computational efficiency}

\begin{figure}[htbp]
\begin{center}
\subfigure[\label{amp}]
{\includegraphics[height=3.0cm,angle=0]{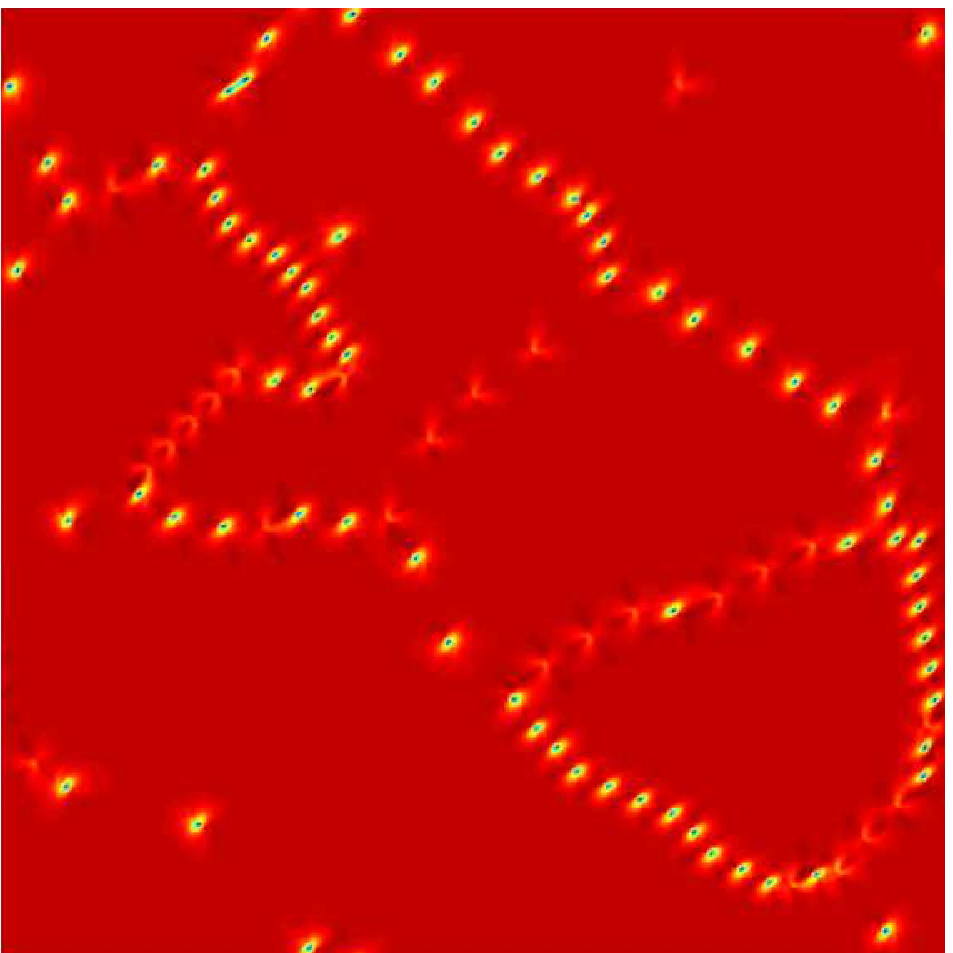}}
\hskip 1cm
\subfigure[\label{phgrad}]
{\includegraphics[height=3.0cm,angle=0]{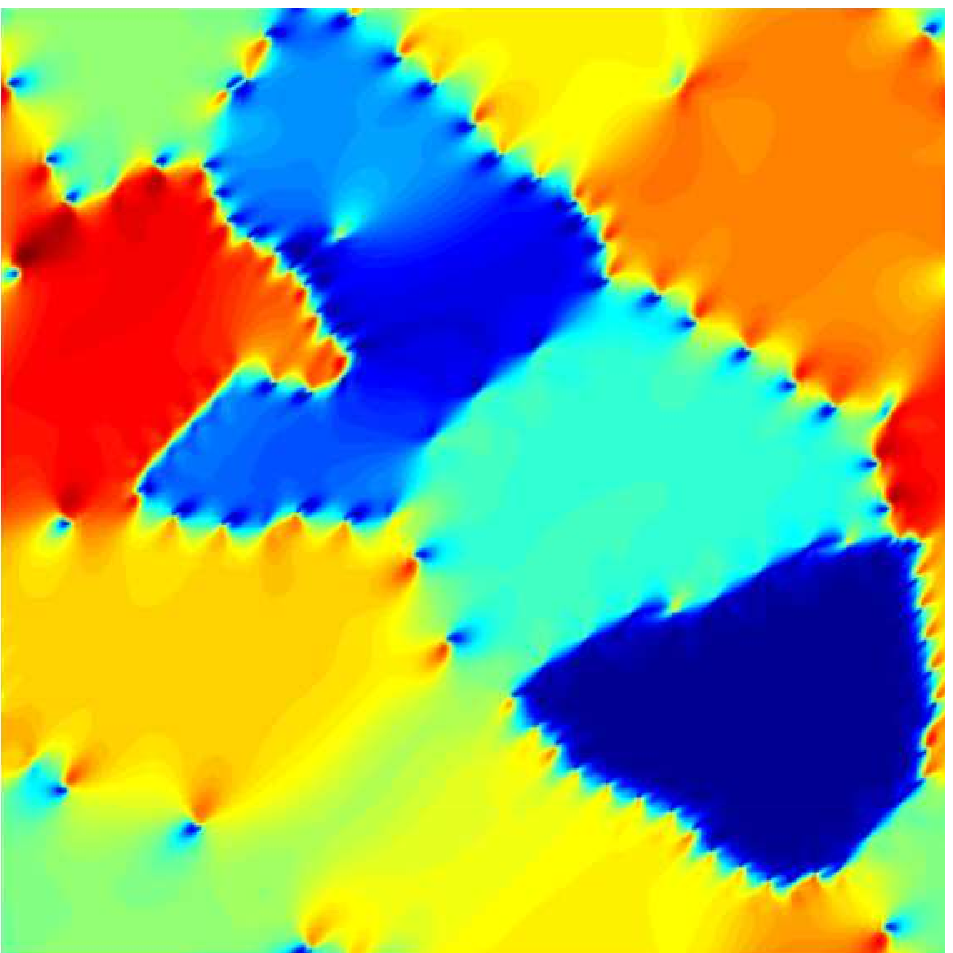}}
\caption{Color map of the spatial variation of the
amplitude (a) and phase gradient (b) of the solution
displayed in Fig.~(\ref{fig:comparison}) at time $t=720.04$.}
\label{fig:amp_phase}
\end{center}
\end{figure}

In order for our scheme to be
computationally efficient, we need to establish that the solutions for
the fundamental mesoscale variables are indeed slowly varying.  In
Fig.~(\ref{fig:amp_phase}) are shown the amplitude and phase gradient of
one of the components during the computation of the two-dimensional
grain growth.  It is evident that the variables are indeed essentially
uniform, except near the edges of the grains.

In order to exploit this property computationally, we need to work with
a formulation which is independent of the particular orientation of our
reference directions.  To this end, we reformulate the RG-PFC equations
for the complex amplitudes $A_j$ etc. in terms of their real amplitude
and phase variables, denoted by $\Psi_j>0$ and $\Phi_j$. Expanding out
the terms in Eq.~(\ref{rgeqn}), and equating real and imaginary parts,
we obtain equations of motion for $\Psi_j$ and $\nabla\Phi_j$ which can
readily be solved by adaptive mesh refinement, to be reported
elsewhere.  This formulation is also important in treating the beats
that can arise if the crystallographic axes of a grain are not
collinear to the basis axes used in the numerical solution.  Such beats
can be dealt with by either adaptive mesh refinement or the polar
coordinate formulation, and will be discussed in detail elsewhere.  As
we have previously shown\cite{PGD}, adaptive mesh refinement algorithms
scale optimally, with the number of computations being proportional to
the number of mesh points at the finest scale, i.e. to the grain
boundary length (two dimensions) or surface area (three dimensions).

%The demonstration calculations in this paper have been performed close
%to the bifurcation point of the PFC equation, but in general the
%analysis must be extended for arbitrary phase diagrams, far from
%threshold.  The appropriate RG equations in the literature\cite{CN, PN,
%NPL, Sasa} for the phase alone break down in the vicinity of defects,
%where the amplitude cannot be treated as slaved to the phase variable,
%and need to be supplemented by the correct amplitude equations in a
%numerically stable way. The PFC model can be extended to include three
%dimensions, multi-component systems, thermal fields and realistic
%atomic correlations.

In summary, we have shown that multiscale modeling of complex
polycrystalline materials microstructure is possible using a
combination of continuum modeling at the nanoscale using the PFC model,
RG and related techniques from spatially-extended dynamical systems
theory.  The PFC model can be extended to include three
dimensions, multi-component systems, thermal fields and realistic
atomic correlations.  Our analysis is ideally-suited
for efficient adaptive mesh refinement, thus enabling realistic
modeling of large-scale materials processing and behavior.

\begin{acknowledgments}

It is a pleasure for the authors to contribute to this special volume
of the Journal of Statistical Physics with an article that so
prominently builds upon earlier work by Jim Langer and Pierre
Hohenberg. NG is delighted to be able to use this opportunity to
express his appreciation to Jim and Pierre for their collaboration and
friendship over the years, especially during the early stages of the
development of the pattern formation field.

This work was supported in part by the National Science Foundation
through grant NSF-DMR-01-21695 and by the National Aeronautics and
Space Administration through grant NAG8-1657.

\end{acknowledgments}

%\bibliographystyle{apsrev}
%\bibliography{./multiscale_langer}

\end{document}